# ATTITUDE CONTROL OF THE ASTEROID ORIGINS SATELLITE 1 (AOSAT 1)

## Raviteja Nallapu,[*] Saumil Shah,[†] Erik Asphaug,[‡] and Jekan Thangavelautham[§]

Exploration of asteroids and small-bodies can provide valuable insight into the origins of the solar system, into the origins of Earth and the origins of the building blocks of life. However, the low-gravity and unknown surface conditions of asteroids presents a daunting challenge for surface exploration, manipulation and for resource processing. This has resulted in the loss of several landers or shortened missions. Fundamental studies are required to obtain better readings of the material surface properties and physical models of these small bodies. The Asteroid Origins Satellite 1 (AOSAT 1) is a CubeSat centrifuge laboratory that spins at up to 4 rpm to simulate the milligravity conditions of sub 1 km asteroids. Such a laboratory will help to de-risk development and testing of landing and resource processing technology for asteroids. Inside the laboratory are crushed meteorites, the remains of asteroids. The laboratory is equipped with cameras and actuators to perform a series of science experiments to better understand material properties and asteroid surface physics. These results will help to improve our physics models of asteroids. The CubeSat has been designed to be low-cost and contains 3-axis magnetorquers and a single reaction-wheel to induce spin. In our work, we first analyze how the attitude control system will de-tumble the spacecraft after deployment. Further analysis has been conducted to analyze the impact and stability of the attitude control system to shifting mass (crushed meteorites) inside the spacecraft as its spinning in its centrifuge mode. These analyses been performed to bound the science payload mass and identify fail-safe methods to guarantee spin stability and stop spinning when commanded to do so. The spacecraft will need to remain stationary when transmitting important science data to Earth and for conducting accretion experiments. AOSAT 1 will be the first in a series of low-cost CubeSat centrifuges that will be launched setting the stage for a larger, permanent, on-orbit centrifuge laboratory for experiments in planetary science, life sciences and manufacturing.

---

[*] PhD Student, Space and Terrestrial Robotic Exploration Laboratory, Arizona State University, 781 E. Terrace Mall, Tempe, AZ.
[†] GNC Engineer, United Launch Alliance (ULA),
[‡] Professor and Ronald Greeley Chair of Planetary Science, Space and Terrestrial Robotic Exploration Laboratory, Arizona State University, 781 E. Terrace Mall, Tempe, AZ
[§] Assistant Professor, Space and Terrestrial Robotic Exploration Laboratory, Arizona State University, 781 E. Terrace Mall, Tempe, AZ

## INTRODUCTION

Missions to asteroids and comets help us answer some of the fundamental questions about origins of the solar system, earth and the building blocks of life. However, these missions present unique challenges owing to their low gravity environments, since getting into an orbit around an asteroid or landing is hard. Techniques like grappling can end up pushing rubble particles, while landings might experience forces enough to attain bounce-off with escape velocities. These problems were observed in Hayabusa -1[1], Phillae[2], and Phobos missions[3]. There is an important need to gain fundamental understanding of asteroid physics, formation and material science to enable ambitious future landing, sample return and resource-mining missions. What better way to prepare for these missions by simulating asteroid surface conditions.

However, simulating asteroid surface conditions remains a formidable challenge. The challenge comes from simulating the low-gravity conditions. Figures 1 present some conventional methods to simulate low-gravity conditions on earth. These include parabolic flight, use of neutral buoyancy within large water tanks and drop towers. In a parabolic flight, an aircraft is maneuvered to create brief periods of micro-gravity conditions last 10-20 seconds[4]. Neutral buoyancy methods suspend objects in water, with buoyant attachments that compensates for the objects mass[5]. Finally, drop towers contain a chamber that fall for 5-10 seconds enabling the contents of the chamber to briefly experience microgravity conditions[6]. These conventional methods simulate low-gravity conditions for too brief a time period or impose simulation artifacts that prevent correlation with real asteroid surface conditions.

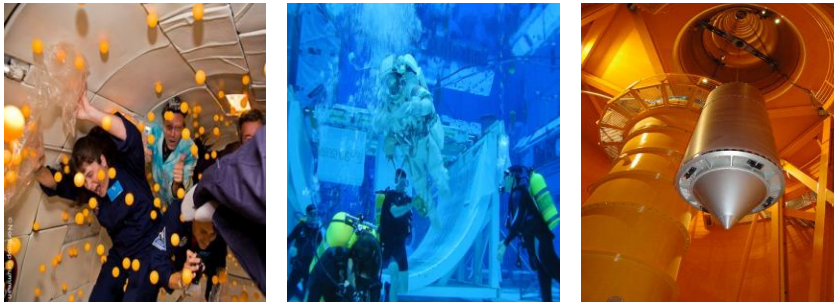

**Figure 1. Methods to simulate low-gravity conditions include use of parabolic flight (left), neutral buoyancy in large water tanks (center) and use of drop towers (right).**

A promising solution to simulating low-gravity conditions is using a centrifuge operating in Low-Earth Orbit[7]. The centrifuge consist of a mass $m$, spinning at a radius $r$, at an angular velocity $\omega$, which create a centrifugal force of magnitude $F_c = m\, r\, \omega^2$. The concept of a space centrifuge is not new and has been a popular topic of science fiction. However, we have yet to see a habitation centrifuge or centrifuge science laboratory operate in space. Our focus is to create a centrifuge science laboratory to simulate the surface conditions and the physics of small bodies.

We have proposed utilizing a 3U CubeSat to test the concept. CubeSats are emerging as low-cost platform to perform space science and technology research. They offer the possibility of short development times, wide use of Commercial-Off-the-Shelf Technologies (COTS), frequent launches and training of graduate and undergraduate students. Our first CubeSat science laboratory mission is called Asteroid Origins Satellite-1 (AOSAT-1)[8,17,18]. The spacecraft, a 3U, 34 cm × 10 cm × 10 cm (size of a loaf of bread) will contain a science chamber that takes two-thirds of the spacecraft volume and contain crushed meteorite. One third of the spacecraft contains the spacecraft electronics, Guidance Navigation and Control (GNC), communications electronics and the power system. GNC plays a critical part on the AOSAT-1 demonstrator mission and is the focus of this paper. This paper discusses the GNC strategies used to develop the AOSAT-1 spacecraft. The spacecraft produces artificial gravity by spinning at 1 RPM as shown in Figure 2, using magnetorquers and a reaction-wheel. This is sufficient to simulate the gravitational forces experienced on a sub 1 km asteroid. A major advantage of a centrifuge science laboratory such as AOSAT 1 is that it can simulate asteroid surface conditions without having to go to an asteroid, which remains a long and expensive endeavor. These centrifuges can help use prepare for future mission by testing new technologies under asteroid conditions.

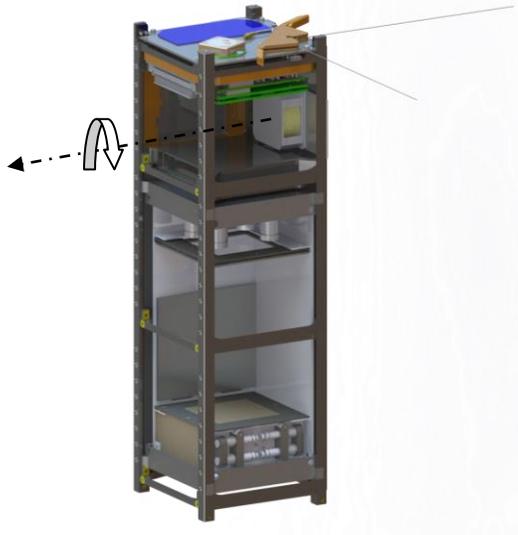

**Figure 2. AOSAT-1 Model with its spin axis.**

We begin with an overview of the mission and concept of operations. We then move to outline attitude control requirements of the mission followed by presentation of the rigid-body equations of motion and the physics. A discussion on attitude determination of the spacecraft with mass uncertainties is presented followed by simulation results of the expected GNC performance. Finally, we conclude the paper with a summary of findings and future work.

**MISSION**

AOSAT-1 will be launched aboard a rocket resupply mission to the International Space Station (ISS). The CubeSat will be deployed from ISS into a low earth orbit (LEO) at a 370-440 km altitude. Each orbit is about 92 minutes long. The spacecraft upon deployment undergoes a mandatory 20 minutes of unpowered flight. The concept of operations of AOSAT-1 is summarized in Figure 3. The spacecraft is expected to tumble due deployment disturbances. Once the spacecraft is powered, it will proceed with a de-tumbling sequence, followed by first contact with ground control. Following first contact, the spacecraft will undergo a commissioning phase for about 1 month followed by Science-1 phase.

After the commissioning phase, the regolith stowed in the chamber is released into the payload chamber and monitored under microgravity using a suite of cameras. Following the extended microgravity experiments, AOSAT-1 will operate in a centrifuge mode, during Science-2 phase. The spacecraft spins at 1 RPM about the body axis for experiments lasting 1-3 hours. The regolith dynamics will be monitored using the onboard cameras. After each experiment, critical data will be communicated back to ground using a UHF link to the ASU ground station.

Having outlined the major phases of the mission, the Attitude Control System (ACS), will have the following 5 modes: De-tumble, nominal mode, spin mode, de-spin mode, and a safe mode. De-tumble is executed upon power-up, post deployment. In the nominal mode, the space-craft maintains a stationary attitude. In the spin mode, the spacecraft operates as a centrifuge. While in safe-mode, the spacecraft operates on low-power, while polling the attitude determination sensors. The attitude-control actuators are initially turned off to isolate anomalies.

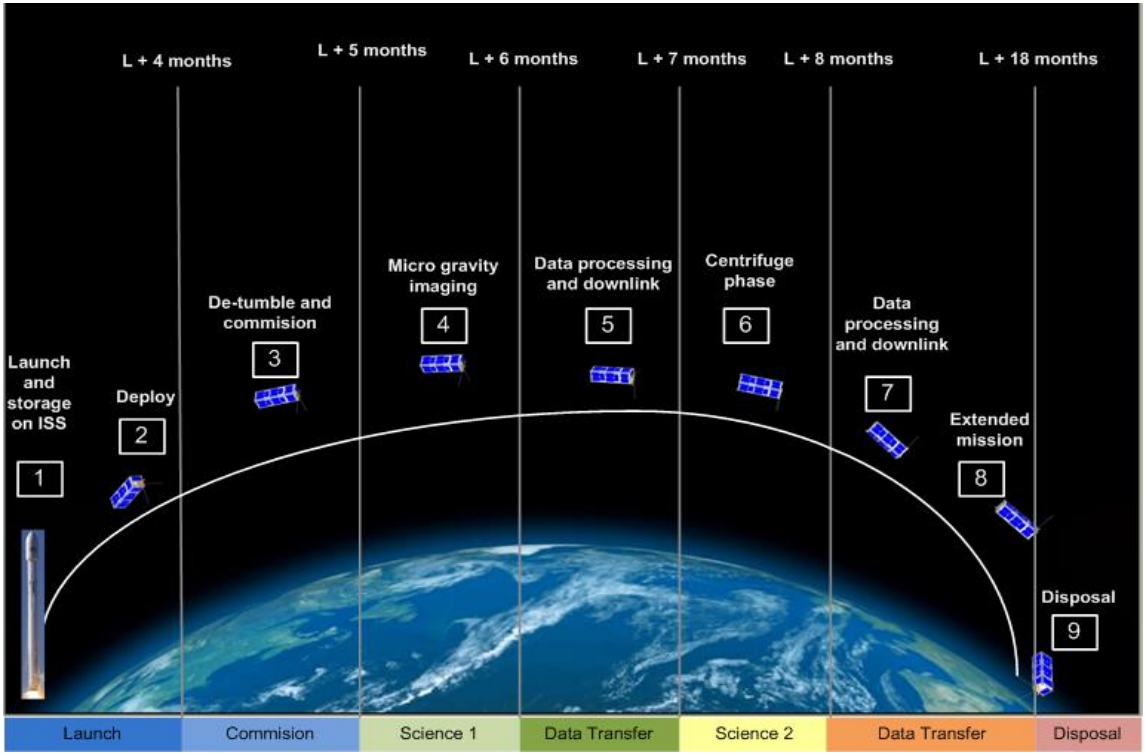

**Figure 3. AOSAT-1 Concept of Operations.**

**ATTITUDE DETERMINATION AND CONTROL SYSTEM (ADCS)**

**Requirements**

A set of performance requirements for AOSAT-1 ADCS system were agreed upon, which are presented in Table 1:

**Table 1. AOSAT-1 ADCS Requirements**

| No: | Requirement |
|---|---|
| 1 | The ADS shall monitor spacecraft angular velocities with an accuracy of less than 0.001 rad/sec |
| 2 | The ADS shall monitor spacecraft attitude with an accuracy of less than 0.001rad |
| 3 | The ADCS shall de-tumble the spacecraft to angular velocities below 0.01 rad/sec within 6 orbits |
| 4 | The ADCS shall stabilize the spacecraft, for communications, with in a 5° half cone about the body z-axis |
| 5 | The ADCS shall spin the spacecraft at 1 RPM about the body x-axis |
| 6 | The ADCS shall spin and de-spin to steady state angular velocity within 1 minute |

Requirements 1 and 2, are the ADS (Attitude Determination System) requirements, and will not be discussed here. Requirements 3-6 are the ACS (Attitude Control System) requirements employed during different phases of operation.

**Subsystem Components**

The AOSAT-1 chassis is composed of TYVAK's Intrepid platform [9]. This platform consists of an Inertial Measurement Unit (IMU), sun sensor, and 3-axis magnetometers and 3-axis magnetorquer coils. In addition, a Blue Canyon micro-reaction wheel is included on the spacecraft to enable higher torques and smooth rotations about the x-axis. Therefore, the total available control input ($\bar{\tau}_c$) is

$$\bar{\tau}_c = \bar{\tau}_m + \begin{bmatrix} \tau_{rw} \\ 0 \\ 0 \end{bmatrix} \qquad (1)$$

Where $\bar{\tau}_m$ denotes the control-torque generated by the magneto-torquer, and $\tau_{rw}$ denotes the control-torque generated by the single reaction wheel along x-axis.

**Attitude Dynamics**

Here we discuss the dynamics model used for attitude determination of the spacecraft. Consider a spacecraft orbiting round the Earth. We define a body frame, $F_b$, and an orbit frame, $F_o$, with the following conventions: The principal axes of the spacecraft will be its basis vectors in $F_b$, with the z-axis being the longest axis, and origin being at the center of mass. In $F_o$, on the other hand, the z-axis points towards center of the Earth, x-axis points toward the direction of the satellites orbital speed, y-axis completes the right-hand triad, as shown in Figure 4.

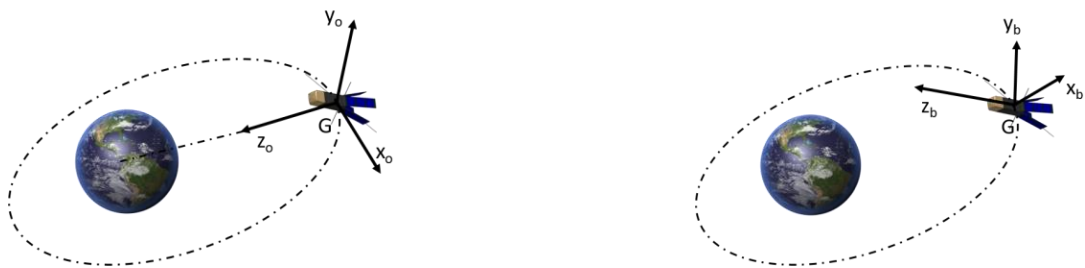

**Figure 4. Cartesian Reference Frames. (Left) Orbit Frame ($F_o$). (Right) Body Frame ($F_b$).**

The orientation of $F_b$ with respect to $F_o$ describes the attitude of the spacecraft. For this work, we use the quaternions to model the attitude of the body frame with respect to the orbit frame. If we let $\omega$ denote the angular velocity of the body frame with respect to the orbit frame, the attitude dynamics are given by the Euler's equations [10, 16] as:

$$\dot{\omega} = -J^{-1}([\omega *]J\omega) + \tau_c + \tau_d \qquad (2)$$

$$\dot{q} = \frac{1}{2}\begin{bmatrix} 0 & -\omega_x & -\omega_y & -\omega_z \\ \omega_x & 0 & \omega_z & -\omega_y \\ \omega_y & -\omega_z & 0 & \omega_x \\ \omega_z & \omega_y & -\omega_x & 0 \end{bmatrix} q \qquad (3)$$

Here, $\bar{\tau}_c$ denotes the control torque as given by equation 2, and $\bar{\tau}_d$ denotes the net disturbance torque. Equations 2 and 3 represent the attitude control model for a spacecraft in 3 dimensions.

**Nominal model**

If we choose *J* to be a constant matrix, and set $\tau_d = 0$, then equations 2 and 3 represents the nominal attitude control model.

**Robust modelling**

On the contrary if *J* can vary, and $\tau_d \neq 0$ then equations 2 and 3 represents the attitude control model for a more robust system. Modelling *J* and $\tau_d$ are discussed as follows:

**Inertia Uncertainty**

Strictly speaking for any real system, *J* cannot be known with perfect accuracy, even if the object studied is rigid and with no moving parts. In such situations, we experimentally determine a nominal value of *J* from experiments, and augment an uncertainty given by the standard deviation of these experiments. However, AOSAT 1, has moving regolith inside it, which constantly changes the moment of inertia compared a simpler rigid body spacecraft. Therefore, we characterize the moment of inertia in a differently using a point mass model.

In this method, various components inside the spacecraft were listed along with their masses $m_i$, and centers of mass inside the body $r_i$. The weighted average of the center of masses of all the components gives the CG of the combined spacecraft.

$$r_g = \frac{\sum_i m_i r_i}{\sum_i r_i} \tag{4}$$

With $r_g$ known, the body frame can now be defined with origin translated to $r_g$, i.e.,

$$r'_i = r_i - r_g \tag{5}$$

Here $r'_i$ denotes the position vector of the $i^{th}$ component in the body frame. Then by defining $\omega = [\omega_1\ \omega_2\ \omega_3]^T$ as a symbolic variable, the angular momentum vector $H = [H_1\ H_2\ H_3]^T$ of the spacecraft while exhibiting pure rotation is given by:

$$H = \sum_i r_i \times m_i (\omega \times r_i) \tag{6}$$

The moment of inertia tensor directly falls out as the co-efficients of $\omega$, given by:

$$J_{ij} = \frac{\partial H_i}{\partial \omega_j} \tag{7}$$

In other words, $J_{ij}$ is given by the co-efficient of $\omega_j$ in the expression for $H_i$. The mass and center of mass catalog in the geometric frame of the chassis for AOSAT 1 is given in Table 2.

**Table 2. Mass Distribution of AOSAT-1**

| Component | Mass (kg) | X-Position (cm) | Y-Position (cm) | Z-Position (cm) |
|---|---|---|---|---|
| Chassis | 1.15 | 0 | 0 | 0 |
| Battery | 0.29 | 0 | 0 | -14 |
| Top & Bottom Panels | 0.07 | 0 | 0 | 0 |
| Side Panels | 0.06 | 0 | 0 | 0 |
| Breakout board-1 | 0.06 | 0 | 0 | -5.1 |
| Daughter board | 0.06 | 0 | 0 | -7.1 |
| Main Computer | 0.06 | 0 | 0 | -7.6 |
| Breakout board-2 | 0.06 | 0 | 0 | -9.6 |
| Power distribution board | 0.06 | 0 | 0 | -11 |
| Camera | 0.21 | 0 | 0 | -2.5 |
| Payload chamber | 0.52 | 0 | 0 | 5 |
| Reaction Wheel | 0.12 | 0 | 0 | -12 |
| Regolith | 0.25 | x | y | z |

The x, y, z co-ordinates of the regolith can be modeled as random variables, within the range of the dimensions of the payload chamber. By placing x, y, z along the edges of the payload chamber, the ranges of CG variation and moment of inertia can be found. These ranges are used to confine the location of x, y, and z.

**Disturbance Torques:**

The typical disturbance torque for a CubeSat when in LEO is: Aerodynamic drag[12], solar radiation pressure[13], and gravity gradient torques[14] that are modeled as follows:

The aerodynamic drag[12] force vector is given by:

$$\overline{F}_{drag} = \frac{-1}{2} C_d \rho (\overline{v}.\overline{v}) \Sigma(\hat{v}.\overline{A_i})\hat{v} \qquad (8)$$

where $C_d$ is the co-efficient of drag, $\rho$ is the density of air, $\overline{v}$ is the translational velocity vector of the spacecraft which has the direction $\hat{v}$, and $\overline{A_i}$ is the area vector of $i^{th}$ leading face of the spacecraft in the direction normal to its face $\hat{A}_i$. Hence the drag torque is given by, when $\overline{r}_{drag}$ and denotes the center of pressure of the drag torque given by the cross-product of the force with the moment arm directed from spacecraft center of mass to center of drag on the leading edge face:

$$\overline{\tau}_{drag} = \overline{r}_{drag} \times \overline{F}_{drag} \qquad (9)$$

The solar radiation pressure[13] force is given by:

$$\overline{F}_{solar} = -\left(\frac{W}{C}\right) \Sigma \left(\overline{A}_i \ (\hat{A}_i.\hat{S})\right)\left[(1-c_{sr})\hat{S} + \left(2c_{sr}(\hat{A}_i.\hat{S}) + \frac{c_{dif}}{3}\right)\hat{A}_i\right] \qquad (10)$$

where W is the solar flux at Earth orbit, C is the speed of light, $\hat{S}_i$ is the direction of the vector joining the Sun to the spacecraft center of mass, and $c_{sr}$ and $c_{dif}$ are the coefficients of specular and diffuse reflection respectively. Hence the solar radiation pressure torque is given by the cross-product of the force with the moment arm directed from spacecraft center of mass to center of solar pressure on the face of the leading edge:

$$\overline{\tau}_{solar} = \overline{r}_{solar} \times \overline{F}_{Solar} \qquad (11)$$

Finally, the gravity gradient torque[14] is given by:

$$\bar{\tau}_{gg} = \frac{3\mu}{|\bar{r}_b|^5}(\bar{r}_b \times J\bar{r}_b) \qquad (12)$$

where, $\bar{r}_b$ is the position vector joining center of earth to the center of mass of the spacecraft. Therefore, the total disturbance torque can now be written as:

$$\bar{\tau}_d = \bar{\tau}_{drag} + \bar{\tau}_{srp} + \bar{\tau}_{gg} \qquad (13)$$

## SIMULATIONS AND RESULTS

Simulations were developed in MATLAB and applied to the 4 functional control modes. The results of the simulation are presented below.

### De-tumble Mode

In this mode, random angular velocities imparted by the CubeSat deployer are eliminated by using magnetorquers alone, ie, $\tau_{rw}=0$. Also, the regolith is stowed in the regolith chamber, and is not allowed to move freely. Therefore, $x$, $y$ and $z$ from Table 2 were chosen as $x=0$, $y=0$, and $z=14$ cm. Consequently, the Inertia Tensor $J$ becomes a constant matrix. A proportional-derivative (PD) control law[15] was applied as follows:

$$\bar{\tau}_c = \bar{\tau}_m = -K_p q_e^* - K_d \bar{\omega}_e \qquad (14)$$

where, $q_e^*$ and $\bar{\omega}_e$ denote the deviations of the present quaternion $q^*$ and angular velocity $\bar{\omega}$ respectively, from their desired values ($q_d^*$, $\bar{\omega}_d$):

$$\bar{\omega}_e = \bar{\omega} - \bar{\omega}_d \qquad (15)$$

$$q_e^* = q^* - q_d^* \qquad (16)$$

Note the $q^*$ is the vector version of quaternion $q$, which includes just the angle dependent terms and ignores the scalar segment. Since we need angular velocities to settle (reach zero) and the spacecraft to align with the orbit axis, we set $q_d^*=0$, and $\bar{\omega}_d=0$. The gains ($K_p$ and $K_d$) are selected by running several simulations with arbitrary positive values, and noting the values that provide the satisfactory performance. To simulate the de-tumble mode, several initial angular velocity vectors were provided on all axis, and the response was obtained by propagating equations 2 and 3 along the orbit. The initial orientation of the spacecraft does not matter, so any orientation can be picked. The gains were chosen as $K_p=$ 9E-5 and $K_d=$ 9E-3. The spacecraft spinning at an unlikely 35 RPM on all 3-axis can be de-tumbled within 6 orbits. The simulated response of the angular velocity and Euler angles is shown in Figure 5 and response times shown in Table 3.

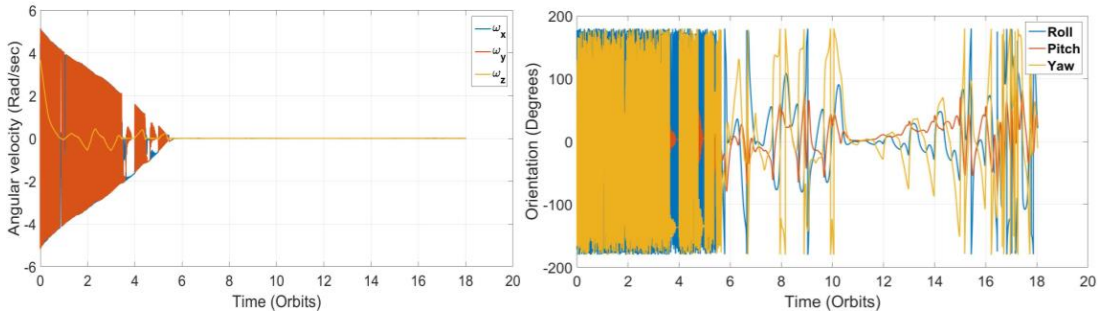

**Figure 5. De-tumbling mode response, with angular velocities (left) and Euler angles (right) shown.**

Note that the body frame does not have to align with the response frame here, since the de-tumble logic places more emphasis on reducing the angular velocity errors than the quaternion errors.

**Table 3. De-tumbling Time**

| Initial Angular Velocity | De-tumbling time (Orbits) |
|---|---|
| 30 | 5.32 |
| 35 | 5.77 |
| 40 | 6.05 |
| 45 | 6.43 |
| 50 | 6.93 |
| 55 | 7.10 |
| 60 | 7.60 |

## Spin Mode

In this mode, the spacecraft is commanded to spin on the body x-axis with 1 RPM, from rest to a user defined angular velocity. The reaction wheel is primarily used for this maneuver. Disturbance torques about the y and z axes can be mitigated by using magnetorquers. The wheel and magnetic control torques, in this case are given by:

$$\tau_{rw} = -K_1 \omega_{xe} \quad (17)$$

$$\bar{\tau}_m = -K_1 \begin{bmatrix} 0 \\ \omega_{ey} \\ \omega_{ez} \end{bmatrix} \quad (18)$$

To simulate this mode, the initial angular velocity was 0 along all 3 axes, and $\bar{\omega}_d$ is set to $[1\ 0\ 0]^T$ RPM in equation 15. Here $\omega_{xe}$, $\omega_{ye}$ and $\omega_{ze}$ are the $x$, $y$ and $z$ components of $\bar{\omega}_e$ respectively.

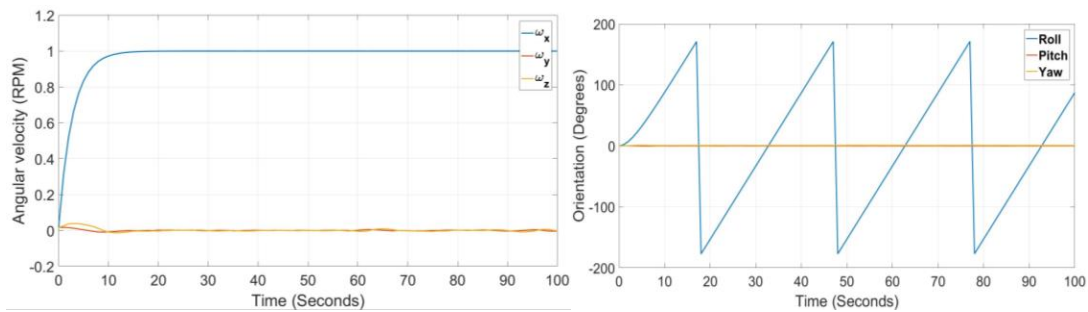

**Figure 6. Spin mode response, with angular velocities (left) and Euler angles (right) shown.**

Gains for this mode we chosen as $K_1$=7E-3 and $K_2$=7E-4. As seen in Figure 6, the spacecraft can achieved a desired, constant angular velocity in less than 10 seconds.

**De-spin Mode**

This mode is exactly the opposite of the spin mode, where we eliminate the user defined spin about the x-axis. Hence, we use the same control logic in equations 17 and 18, but $\bar{\omega}_d$ now is set to $[0\ 0\ 0]^T$ RPM, and initial speed is set to $[1\ 0\ 0]^T$ RPM.

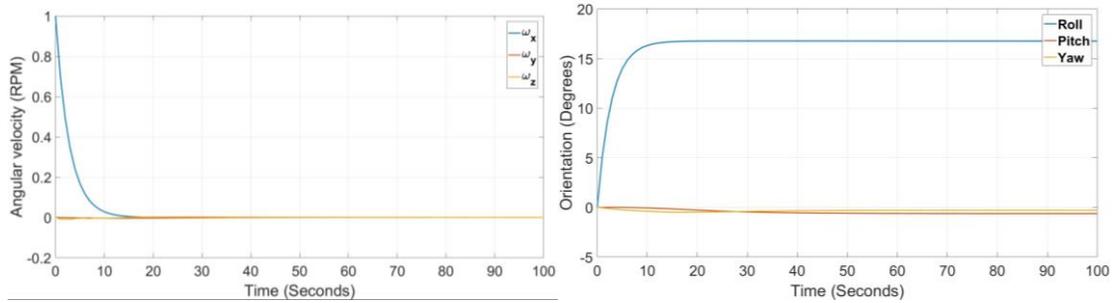

**Figure 7. De-Spin mode response, with angular velocities (left) and Euler angles (right) shown.**

The de-spin mode has a similar settling time response as the spin mode. It is to be noted that the de-spin mode does not align the spacecraft with the orbit frame. However, this mode is immediately followed by the nominal mode (explain below) which does the alignment.

**Nominal mode**

In this mode, the moment of inertia is varied randomly as discussed in the previous section, and the spacecraft is commanded to align its body frame with the orbit frame. This maneuver is made with magnetorquers. The control torque is same as that given by equation 14, however the difference between the 2 modes is that in the de-tumble mode, it is important that the $\bar{\omega}_e = 0$, while the emphasis of the nominal mode is that $q_e^*$=0. To simulate this, the spacecraft was assumed to have 0 initial angular velocity about all axis, however, the initial orientation was π/2 radians (90 degrees) off from the orbit frame on all 3-axes.

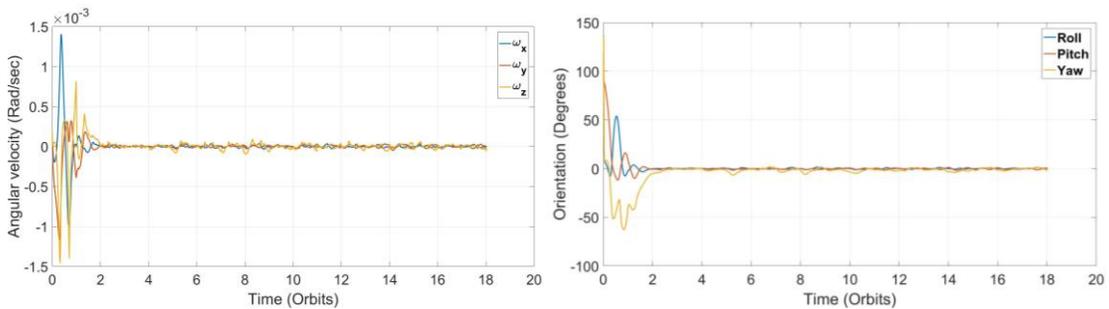

**Figure 8. Nominal mode response, with angular velocities (left) and Euler angles (right) shown.**

The gains chosen were $K_p$=9E-5 and $K_d$=9E-3. In a period of about 2 orbits, the spacecraft tracks the orbit frame as seen in Figure 8. As expected the low output torque from the magnetorquers causes the slow stabilization.

## CONCLUSION

This paper presents the attitude control strategies for AOSAT 1, which will be the first CubeSat centrifuge science laboratory. The concept of operations is presented, which makes it possible to identify the various operational modes of AOSAT 1. The attitude dynamics of the spacecraft, which contains a moving payload of meteorite particles (regolith) is presented. In this paper, we develop the control laws for detumbling, spin-up and spin-down and simulate the results for expected disturbance torques. The simulation results show that the spacecraft can detumble under extreme conditions of 35 RPM. Furthermore, the spacecraft can spin-up and spin-down within 10 seconds under nominal conditions. These results suggest that a CubeSat based low-speed centrifuge science laboratory containing moving payload of few hundred grams is feasible from the point of view of Guidance, Navigation and Control.